%% file: main.tex
\documentclass[10pt,conference]{IEEEtran}

\usepackage{graphicx}

\usepackage{acro}
\usepackage{booktabs}
\usepackage[enable]{easy-todo}
\usepackage{mathtools}
\usepackage[newfloat=true]{minted}
\usepackage[style=default]{caption}
\usepackage{subcaption}
\usepackage{tikz}
\usepackage[normalem]{ulem}
\usepackage{outlines}

\usepackage{stfloats}
\usepackage{svg}
\usepackage{braket}
\usepackage{caption}
\usepackage{subcaption}

\input{acro}

\title{CQM: Cyclic Qubit Mappings}

\author{\IEEEauthorblockN{Maxwell Poster}
\IEEEauthorblockA{Department of Electrical and\\Computer Engineering\\
University of Texas at Austin\\
mposter@utexas.edu}
\and
\IEEEauthorblockN{Sayam Sethi}
\IEEEauthorblockA{Department of Electrical and\\Computer Engineering\\
University of Texas at Austin\\
sayams@utexas.edu}
\and
\IEEEauthorblockN{Jonathan M. Baker}
\IEEEauthorblockA{Department of Electrical and\\Computer Engineering\\
University of Texas at Austin\\
jonathan.baker@austin.utexas.edu}
}

\begin{document}
\maketitle
\thispagestyle{empty}
\begin{abstract}
\input{content/00_abstract}
\end{abstract}

\input{content/10_introduction}

\input{content/30_background}

\input{content/60_conclusion}

\nocite{matplotlib}
\bibliographystyle{plain}
\bibliography{ref}

\end{document}

%% file: acro.tex
\DeclareAcronym{mu}{short=MU, long=machine units}
\DeclareAcronym{dag}{short=DAG, long=directed acyclic graph}
\DeclareAcronym{artiq}{short=ARTIQ, long=advanced real-time infrastructure for quantum physics}
\DeclareAcronym{nisq}{short=NISQ, long=noisy intermediate-scale quantum}
\DeclareAcronym{dsl}{short=DSL, long=domain-specific language}
\DeclareAcronym{jit}{short=JIT, long=just-in-time}
\DeclareAcronym{rb}{short=RB, long=randomized benchmarking}
\DeclareAcronym{dds}{short=DDS, long=direct digital synthesis}
\DeclareAcronym{dac}{short=DAC, long=digital-to-analog converter}
\DeclareAcronym{adc}{short=ADC, long=analog-to-digital converter}
\DeclareAcronym{awg}{short=AWG, long=arbitrary waveform generator}
\DeclareAcronym{afg}{short=AFG, long=arbitrary function generator}
\DeclareAcronym{fpga}{short=FPGA, long=field-programmable gate array}
\DeclareAcronym{yb171}{short=${}^{171}$Yb$^+$, long=Ytterbium 171}
\DeclareAcronym{pmt}{short=PMT, long=photomultiplier tube}
\DeclareAcronym{api}{short=API, long=application programming interface}
\DeclareAcronym{isa}{short=ISA, long=instruction set architecture}
\DeclareAcronym{rfsoc}{short=RFSoC, long=radio frequency system-on-chip}
\DeclareAcronym{rpc}{short=RPC, long=remote procedure call}
\DeclareAcronym{mw}{short=MW, long=microwave}
\DeclareAcronym{bb1}{short=BB1, long=broadband}
\DeclareAcronym{sk1}{short=SK1, long=Solovay-Kitaev}
\DeclareAcronym{spam}{short=SPAM, long=state preparation and measurement}
\DeclareAcronym{cw}{short=CW, long=continuous wave}
\DeclareAcronym{rtio}{short=RTIO, long=real-time I/O}
\DeclareAcronym{sqst}{short=SQST, long=single-qubit state tomography}
\DeclareAcronym{gst}{short=GST, long=gate set tomography}
\DeclareAcronym{1d}{short=1D, long=one-dimensional}
\DeclareAcronym{2d}{short=2D, long=two-dimensional}
\DeclareAcronym{ddb}{short=DDB, long=device database}
\DeclareAcronym{ast}{short=AST, long=abstract syntax tree}
\DeclareAcronym{rf}{short=RF, long=radio frequency}
\DeclareAcronym{mro}{short=MRO, long=method resolution order}
\DeclareAcronym{ir}{short=IR, long=intermediate representation}
\DeclareAcronym{vcd}{short=VCD, long=value change dump}
\DeclareAcronym{gpu}{short=GPU, long=graphics processing unit}
\DeclareAcronym{vliw}{short=VLIW, long=very long instruction word}
\DeclareAcronym{simd}{short=SIMD, long=single instruction multiple data}
\DeclareAcronym{hll}{short=HLL, long=high-level language}
\DeclareAcronym{ci}{short=CI, long=continuous integration}
\DeclareAcronym{hdl}{short=HDL, long=hardware description language}
\DeclareAcronym{mems}{short=MEMS, long=microelectromechanical systems}

\DeclareAcronym{dax}{short=DAX, long=Duke ARTIQ extensions}
\DeclareAcronym{staq}{short=STAQ, long=software-tailored architecture for quantum co-design}
\DeclareAcronym{rc}{short=RC, long=red chamber}

%% file: content/00_abstract.tex
Quantum computers show promise to solve select problems otherwise intractable on classical computers. However, noisy intermediate-scale quantum (NISQ) era devices are currently prone to various sources of error. Quantum error correction (QEC) shows promise as a path towards fault tolerant quantum computing. Surface codes, in particular, have become ubiquitous throughout literature for their efficacy as a quantum error correcting code, and can execute quantum circuits via lattice surgery operations. Lattice surgery also allows for logical qubits to maneuver around the architecture, if there is space for it. Hardware used for near-term demonstrations have both spatially and temporally varying error results in logical qubits. By maneuvering logical qubits around the topology, an average logical error rate (LER) can be enforced. We propose cyclic qubit mappings (CQM), a dynamic remapping technique implemented during compilation to mitigate hardware heterogeneity by expanding and contracting logical qubits. In addition to LER averaging, CQM shows initial promise given it's minimal execution time overhead and effective resource utilization.

%% file: content/10_introduction.tex
\section{Introduction}

Because of the noise inherent in quantum computing processes, large-scale quantum computation requires quantum error correction (QEC). One of the most popular choices is the surface code \cite{fowler2012surface} for which logical qubits are $d \times d$ patches of physical qubits. For program execution, there are several models including lattice surgery and braiding. While neither is definitively superior, a large portion of architecture studies have focused on the prior \cite{brown2017poking,horsman2012surface}. 

For NISQ applications, program qubits must be mapped to hardware qubits and additional communication operations, typically in the form of SWAP gates, must be inserted to interact arbitrary pairs of program qubits. The same is true for programs mapped to a surface code architecture, except program qubits are mapped to \textit{patches} corresponding to an error corrected qubit, and in the case of this paper to a surface code tile \cite{litinski2019game}. Unlike in NISQ, however, communication operations do not necessarily modify the program hardware qubit mapping and instead long-range operations can be interacted in constant depth, regardless of their distance, so long as there is a contiguous channel of logical \textit{ancilla} between interacting logical qubits.

For this reason, prior works on surface code architectures have focused on \textit{static} mappings which do not change, change very little, or always occupy the same set of tiles over the course of the program's execution \cite{litinski2019game, hua2021autobraid, zhu2024ecmas, molavi2023compilation}. This is typically a fair assumption as even when there are routing conflicts because of shared resources, delays are usually at most a small constant number of stall cycles. Some prior work has added some dynamic remapping via SWAPs in order to maximize parallelism and minimize congestion, but does so strictly for congestion reduction \cite{hua2021autobraid}.

Similarly, prior work almost exclusively assumes that physical error rates in error corrected systems are both spatially and temporally constant. That is, if $p_L(i, t)$ is the logical error rate of physical location $i$ at time $t$ then $p_L(i, u) = p_L(j, v)$ for every $i, j, u, v$. Some prior work has explored situations with burst errors \cite{suzuki2022q3de} or with accumulating leakage errors \cite{vittal2023eraser} but provide solutions which are either too specific (e.g. leaking reduction circuits only), high overhead (frequent burst detection), or practically infeasible (excessive rollbacks to recompute corrections). Most importantly, nearly all prior work on error-aware compilation for physical or fault-tolerant computing requires knowledge (a priori or gleaned) about the current (or past) state of the machine. Such information about hardware error rates is likely unreliable, for example gathered too far in the past or predicted from low shot counts during computation and so mapping decisions are made with at least partially inaccurate information.

In this work, we propose a dynamic remapping technique which reshuffles the positions of logical qubits on device as frequently as possible to \textit{guarantee} an average case performance by ensuring that the logical error rate of every program qubit is the average logical error rate of occupied positions. Our technique simultaneously enables frequent physical qubit resets to flush leakage as well as averages out or avoids burst error rates by relocating logical qubits periodically with minimal overhead. We propose architectural designs to support movement without high compilation or runtime overheads. Movement is of course affected by ancilla availability and program features, e.g. qubit idling time so as to not delay total program execution times. 

\begin{figure*}
    \centering
    \scalebox{0.8}{
    \includegraphics[width=\textwidth]{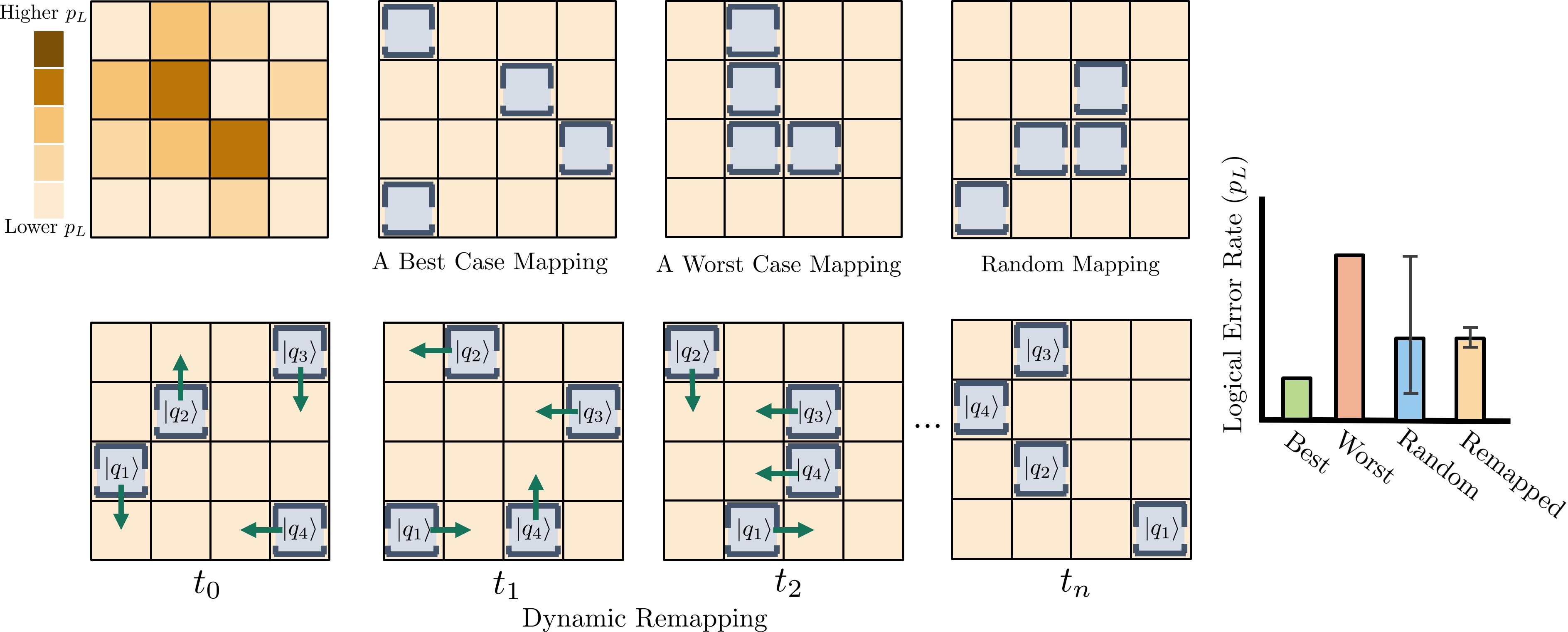}}   \caption{Current quantum hardware systems have qubits with non-uniform error rates which lead to logical tiles with variable logical error rates $p_L$ (top left). Because of the long-running nature of error corrected programs (hours to days), the logical error rate of a given tile is likely unknown. Compilers which use static mappings (top) can result in placing qubits in more or less favorable positions, even though the errors may not be known but cannot provide a guarantee. The alternative (bottom) and our proposal is to have continuously changing mappings so that no program qubit is mapped to the same position for too long and instead we can guarantee the average case logical error rate for every tile.}
    \label{fig:introfig}
\end{figure*}

In Figure \ref{fig:introfig} we give a high level view of generic static mapping strategies versus a high level view of our proposal. For a given surface code, the 4$\times$4 patch of colored squares,  architecture with various logical error rates (darker colors indicate approximately worse logical error rates) spread across the device, there are clearly some mappings of qubits to the device which minimize the total error per qubit. However, the actual error rates of positions on the board cannot be tracked easily. Thus the same mapping algorithm may choose sub-optimal mappings, such as all high-error tiles or some mixture. This is indicated by all mappings being shown on a uniformly colored grid when the user has uncertainty about the exact errors of the device. With space available, our alternative proposal (the second line) shuffles the mapping over time when data qubits are idle. 

The primary contributions of this work are:
\begin{itemize}
    \item We demonstrate spatial and temporal variation can cause uncertainty in the logical error rate of surface code qubits which can cause static compilation techniques to inadvertently choose worst-case mappings of program qubits to hardware qubits.
    \item We propose dynamic remapping of logical qubits during runtime to mitigate quantum system failures (e.g. burst errors, leakage, and drift) by enforcing an average logical error rate on each logical qubit. 
    \item We propose cyclic qubit mappings (CQM), a compilation strategy that dynamically remaps qubits during runtime promising minimal execution time overhead, and anticipate improved execution times in select cases. Our movement strategies have minimal cycle overhead while also ensuring dynamic remapping.
\end{itemize}

%% file: content/30_background.tex
\section{Background}
\label{sec:background}

\begin{figure*}
    \centering
    \scalebox{0.75}{
    \includegraphics[width=\textwidth]{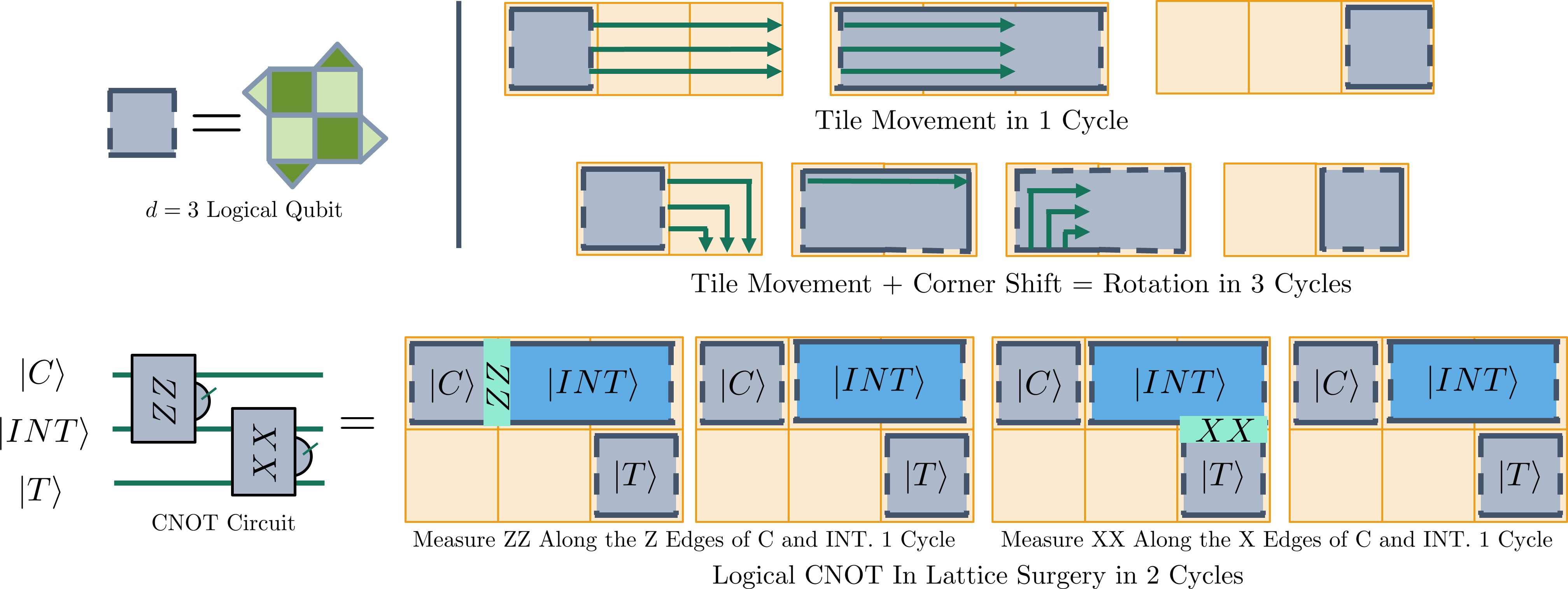}}
    \caption{For this work we abstract away the physical requirements of all tiles (top left) so that every Surface Code tile is represented as a single colored square with 2 types of boundaries: Z (dashed) and X (solid). Tiles can be moved (top middle) by expending the patch into available ancillary space and then measuring old space. This can be done in a single code cycle. The boundaries can also be modified by shifting corners using adjacent ancillary space and expansion. This can be done in 3 cycles. We also utilize a lattice surgery version of CNOT gates for multi-qubit gates (bottom). To do so we measure ZZ between the control and a prepared intermediate (INT) ancilla and then measure XX between INT and the target. This can be done in 2 cycles. }
    \label{fig:background}
\end{figure*}

\subsection{Quantum Error Correction and Surface Codes}
In order to protect against physical errors, physical qubits are encoded into \textit{logical qubits}. Errors can be detected and corrected by measuring syndromes via \textit{syndrome extraction circuits} which record parity information about the physical qubits without destroying the underlying superposition required to maintain the quantum state. For this work, the specific mechanics of these operations is not necessary and we refer the reader to \cite{fowler2012surface, litinski2019game}. The actual execution time of programs will depend on the exact implementation of these circuits, but it is typical to consider \textit{code cycles} as a timing abstraction and record operations in terms of code cycles. Some logical operations (those which transform the logical state as opposed to the physical state) can be performed transversally by applying a set of local \textit{physical} operations. Unfortunately, no quantum code supports a universal set of transversal operations and requires alternatives such as magic state distillation or code switching \cite{gottesman1998theory, kubica2018abcs, bravyi2012magic}.

One of the most popular and promising codes in the near term is the Surface Code (more precisely the rotated surface code). In Figure \ref{fig:introfig} (top left) we represent logical qubits as square tiles with two types of boundaries, $X$ and $Z$. Surface codes are appealing due to their low hardware requirements, needing low weight and physically local parity check circuits; each small light (dark) green square corresponds to an X (Z) parity check. They also have high thresholds and well-studied decoders such as Union-Find \cite{delfosse2021almost} and Minimum-Weight Perfect Matching (MWPM) \cite{higgott2022pymatching}, which can be made hardware efficient \cite{vittal2023astrea}.

Surface codes have well defined logical operators via braiding or lattice surgery. In this work we will focus on lattice surgery operations which can be found in \cite{litinski2019game}. The most important operation is tile movement, which takes $1$ code cycle via tile deformation and long-distance multi-qubit operations, e.g. CNOT performed by preparing an intermediate state $\ket{INT} = \ket{0}$ and measuring the ZZ operator along the Z boundary of the control $\ket{C}$ and $\ket{INT}$ and then measuring the $XX$ operator along the X boundary of $\ket{INT}$ and $\ket{T}$. It is possible that because of how qubits are mapped (see below) that the required edges are not exposed to adjacent ancilla. We summarize these operations in Figure \ref{fig:background}. Notice all multi-qubit operations in this architecture require mediating ancilla qubits of at least equal height/width as $d$.

Most importantly for this work is that surface code architectures are easy to construct and reason about. A surface code architecture is a large fabric of $d\times d$ patches, where $d$ is the code distance. The architecture must have enough space to support all $N$ program qubits and a sufficient number of ancilla, $M$, so that between any two interacting data qubits there exists a path of ancilla. The choice of $d$ depends on the requisite logical error rate of the application, $p_{L, targ}$. This depends on both the physical error rate $p_{phys}$ and $d$, i.e. $p_{L} = f(p_{phys}, d)$. 

With endless quantum resources, we could always choose a $d' >> d_{targ}$ which exhibits a quadratic increase in the physical device requirement per logical qubit. For any drift, leakage, or infrequent burst errors, choosing a larger distance ``solves'' the issue at the cost of severe overhead. We expect quantum resources to be limited and should consider approaches to mitigate variance beyond modifying $d$.

\subsection{Hardware Variance, Time-Varying Errors and Leakage}
Quantum hardware developers have made great strides in improving the quantity, reliability, and consistency of available qubits across several potential technologies. Despite the rapid improvements to the underlying hardware, devices still exhibit significant variance in their error rates both spatially and temporally \cite{carroll2024subsystem, Qiskit, suzuki2022q3de}. Error rates may drift slowly or experience transient burst errors \cite{zhang2023disq, ravi2023navigating} in part due to the presence of two-level systems (TLS) defects \cite{klimov2018fluctuations, de2020two}. Defective qubits, even for a short time, can result in significant changes in the \textit{logical} error rate of codes using them. Non-uniform error rates in the physical hardware yields non-uniform error rates in the grid of logical qubits. We give an explicit example of this case in the Section \ref{sec:motivation}.

The physical error rate of hardware is determined via characterization and calibration methods, e.g. randomized benchmarking \cite{knill2008randomized}. These procedures require the qubits to be measured repeatedly and therefore cannot be done during the program execution. Some changes in physical error rate can be detected within some certainty \cite{suzuki2022q3de}, but it's generally expected that the true logical error rates of the system are \textit{unknown} throughout computation  \cite{gidney2021factor}. An initial set of the best qubits are not always guaranteed to be the best later on.

Finally, physical qubits are defined on two logical states $\ket{0}$ and $\ket{1}$. However, many physical systems have other states e.g. $\ket{2}, \ket{3}$ and so on. Normal operation of the qubit can result in some population being \textit{leaked} into these higher states which is undesirable. Leakage can be reduced or even eliminated by physically swapping the physical state into an ancilla $\ket{0}$, resetting the original qubit with a measurement, and then swapping back. In this work, as we will see, the same effect can be achieved if the entire logical qubit is relocated and the entire original patch of qubits is measured and reset. Some prior work has explored both the systematic detection and use of so-called ``Leakage Reduction Circuits'' \cite{vittal2023eraser}. 
\subsection{Compilation to QEC Systems}
\begin{figure}
    \centering
    \includegraphics[width=0.9\columnwidth]{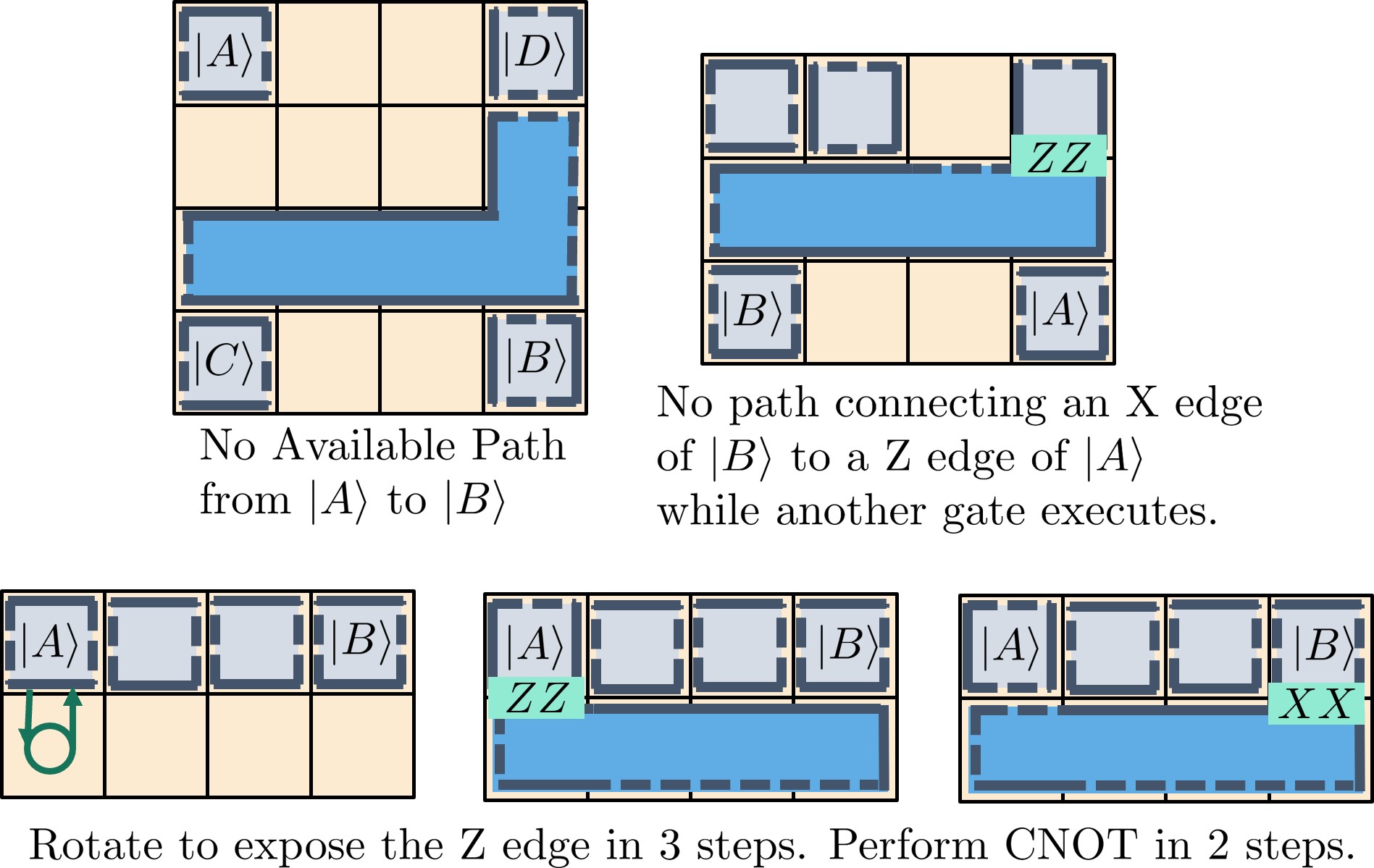}
    \caption{Constraints on how operations can be parallelized. In the top half the objective is to in parallel execute a CNOT between A and B and C and D. If the routing path in blue is already allocated for the CNOT between C and D then we must stall before executing the CNOT between A and B. This can also be true even if there is a path between A and B (top right) but not between the proper edge types. In the bottom, we sometimes are forced to perform rotations of the qubits rather than stalling for availability when qubits are packed densely.}
    \label{fig:routing_constraints}
\end{figure}
Beyond circuit optimization, for current physical quantum programs the compilation problem involves three main steps. First is construction a mapping of program qubits to hardware qubits. Specifically, we construct an injective map $\varphi: A \rightarrow H$ which assigns a hardware qubit $h$ to each program qubit $a$. Hardware qubits have limited connectivity, indicated by the hardware interaction graph $G_H$ which determines which pairs of hardware qubits can interact, i.e. hardware qubits $h_i, h_j$ can interact only if $(h_i, h_j) \in E(G_H)$. Second is the addition of routing operations, such as SWAPs or MOVEs, so that when $(\varphi(a_i), \varphi(a_j)) \not\in E(G_H)$ we can modify the mapping $\varphi$. Instead of a single static mapping, we instead consider a sequence of mappings at each time step $t$, $\varphi(t)$. Finally, operations are scheduled such that execution time is minimized. For NISQ hardware, the cost of routing scales linearly with distance between interacting qubits both in time and gate count. As such, most strategies involve some form of shortest path finding.

Compilation for surface code architectures are similar in the sense that long-range operations require special actions. As suggested in Figure \ref{fig:routing_constraints}, there are two requirements for long-range interactions: 1) a contiguous channel of ancilla between the control and target and 2) a Z edge of the control must be adjacent to a Z edge of the intermediate ancilla and the X edge of the target must be adjacent to the X edge of the intermediate ancilla. When condition 1) is violated, we typically need only stall while other operations using the ancilla complete. When condition 2) is violated, we typically need to insert an edge rotation on either or both of the control and target. 

Magic states, otherwise known as resource states, are crucial for executing arbitrary $Z$ rotations in a surface code architecture. Magic state \textit{factories} prepare a $T$ gate, which is used to induce a phase. $T$, in combination with $S$ and $H$ gates, can approximate $Z$ rotations up to arbitrary thresholds \cite{ross2014optimal}. The previously mentioned constraints apply for the injection of magic states via gate teleportation \cite{gottesman1999quantum}.

\begin{figure}
    \centering
    \includegraphics[width=\linewidth]{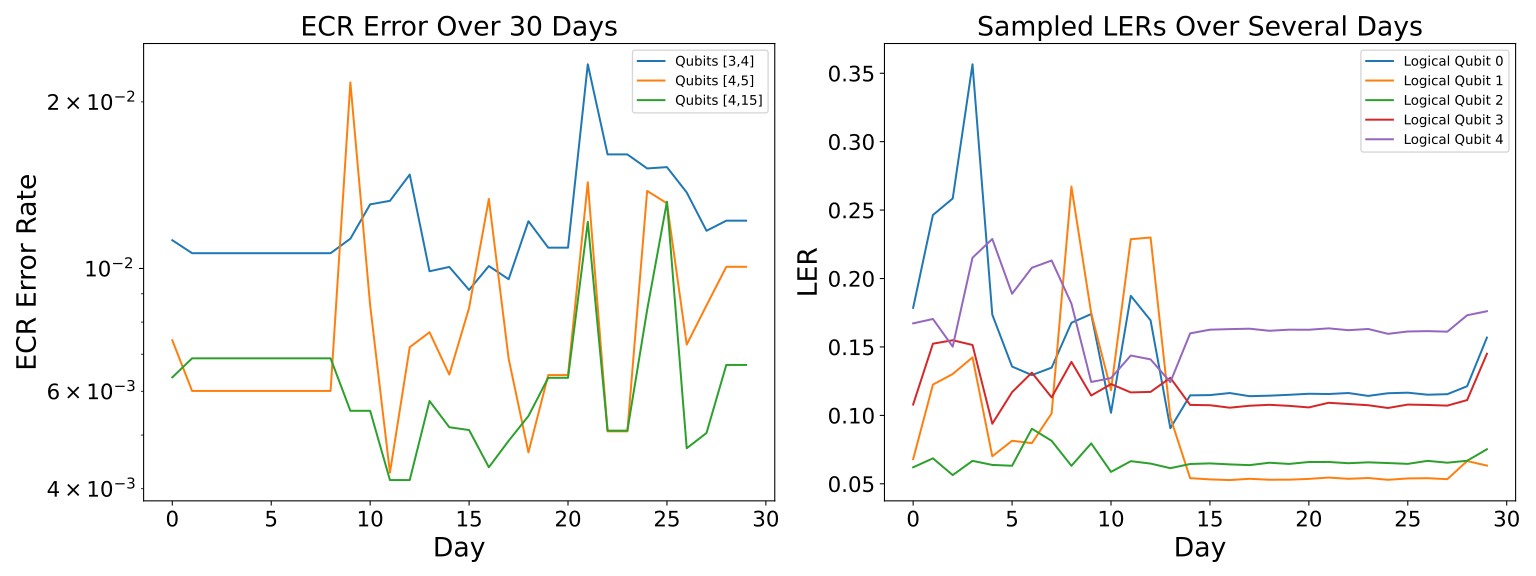}
    \caption{Left: ECR (Error per Clifford Rate) tracked on the IBM Kyoto machine for 30 days. Right: Logical error rate (LER) computed with Google's Stim simulator \cite{Gidney_2021} using physical error rates sampled from the IBM Kyoto machine over 30 days.}
    \label{fig:ecr_30}
\end{figure}
\section{Motivation: Uncertainty in Logical Error}\label{sec:motivation}
In this section, we give an explicit example of how physical error rates can produce both spatial and temporal uncertainty in the logical error rate of qubits which use them. In this demonstration, we query IBM hardware \cite{Qiskit} for physical error rates on each of their qubits and links every day for approximately 30 days, which they report about once a day. The time between calibrations is subject to even further variance, but for simplicity we do not give a continuous view of logical error rate changes. To generate the logical error rates, because the hardware of IBM machines is not 2D grid, we simply assign the the error rates from $q_0$ on hardware to $q_0$ in the $d=3$ surface code tile. 
We can see the fluctuation of the physical error rates for a select number of qubits from the IBM Kyoto device in Figure \ref{fig:ecr_30}, as well as the resulting logical error rate over time. Many of the logical error rates seen in Figure \ref{fig:ecr_30} occur above the fault-tolerant threshold, due to in part to the ECR gate error rates, as well as other sources of physical error such as single gate error, readout errors, etc.

\section{Average Case}
\begin{figure}[b]
    \centering
    \includegraphics[width=0.9\columnwidth]{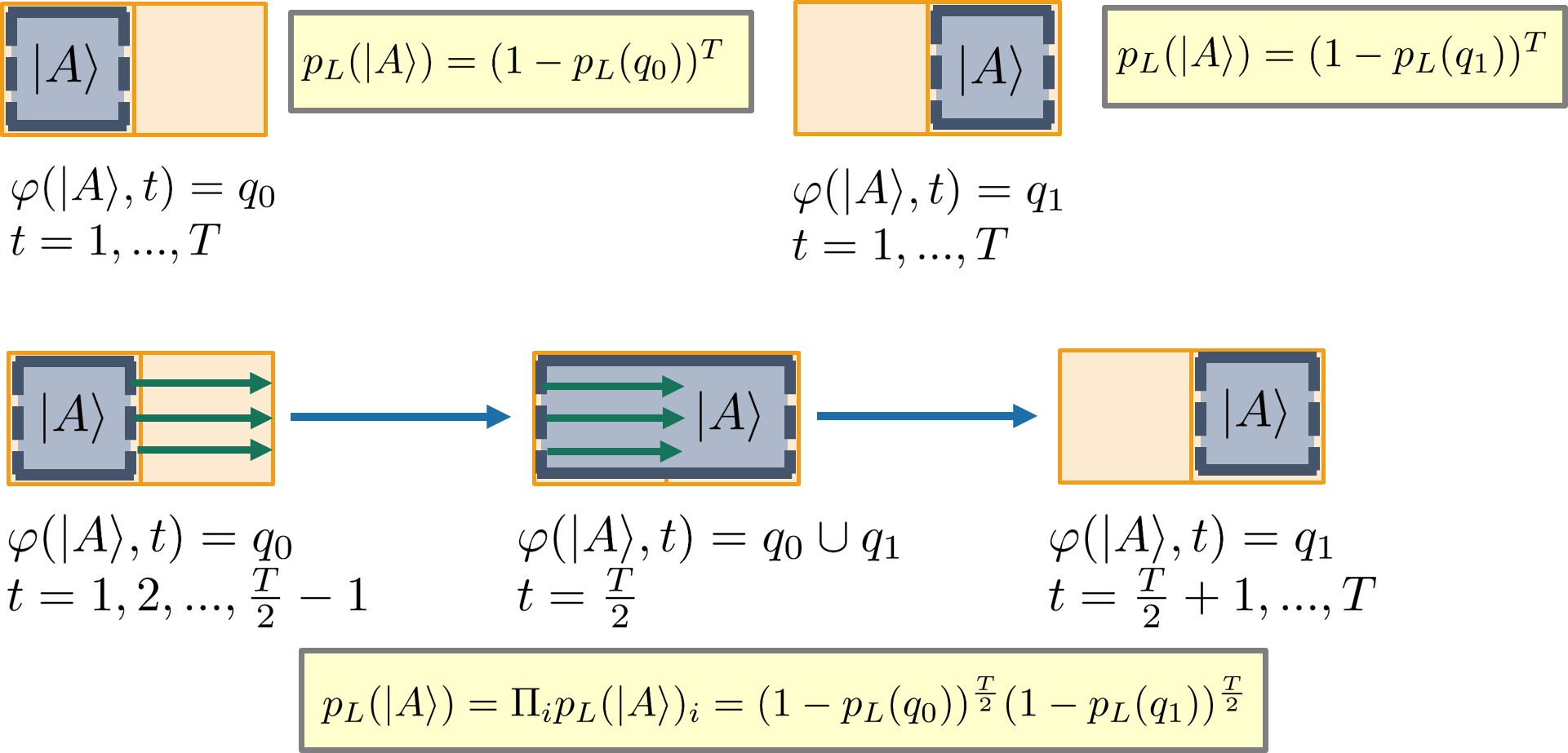}
    \caption{In this proof of concept with a $2\times 1$ sized patch and a single logical qubit we have three options. The first is to keep A in either position 1 or in position 2 and when error rates are unknown mid computation, so too is the logical error rate (top). In our proposal, we simply modify the mapping of A throughout the program so that it spends equal time in every position and so gets the average case logical error rate guaranteed (bottom).}
    \label{fig:enter-label}
\end{figure}
In this section, we give proof of concept examples for how dynamic remapping can result in an approximate \textit{guarantee} about the logical error rates of logical qubits in light of hardware uncertainty. 

\subsubsection{Example 1: One Logical Qubit with 2 Positions} For the simplest example, consider a $2\times 1$ grid with positions $q_{0}$ and $q_{1}$ and single logical qubit $\ket{A}$ to be mapped. For simplicity, assume $p_L(q_{i})$ gives the logical error rate of position $q_{i}$ and without loss of generality assume $p_L(q_{0}) < p_L(q_{1})$. For a program which runs for $T$ code cycles, the estimated probability of success is given as $(1 - p_L)^T$. 

There are 3 primary scenarios: 1. $\ket{A}$ is mapped to $q_0$ for all $T$ cycles, i.e. $\varphi(\ket{A}, i) = q_0, t = 1, ..., T$ 2. $\varphi(\ket{A}, i) = q_1, t=1, ..., T$ or the average case scenario when half way through we modify the mapping so that $\varphi(\ket{A}, t) = q_0$ for $1 \le t \le \frac{T}{2} - 1$ and $\varphi(\ket{A}, t) = q_1$ otherwise. Notably, the movement requires the qubit to be mapped to \textit{both} of the positions for a single cycle, i.e. $\varphi(\ket{A}) = q_0 \cup q_1$, so it should obtain the worst-cast error of both positions, assuming equivalent X and Z type error rates (as we will do for simplicity). This suggests that \textit{infrequent} movement is preferred. Consider the case where every other cycle we expand $\ket{A}$ into both positions, then it will take the worst case error in 75\% of cycles which is non-ideal. However, by moving once halfway through the idle time, we ensure the equal 50\% occupancy. 

How does this affect the logical error rates? In scenario 1 (2) $\ket{A}$ see's the logical error rate of $q_0$ ($q_1$) and so has a single fixed probability of success $(1 - p_L(q_0))^T$ ($(1 - p_L(q_1))^T$). In the mixed version, we see the error rate of $q_0$ for half the time and the error rate of $q_1$ for half the time giving the next probability of success as $(1 - p_L(q_0))^\frac{T}{2}(1 - p_L(q_1))^\frac{T}{2}$. This is less optimal than the best case scenario but it's certainly better than the worst case scenario. When error rates are not confidently known, this ensures an average case error rate rather than risking the worst case. We can also obtain an average case by running the program many times with randomized mappings but this causes excessive overheads which we avoid.

\subsubsection{Example 2: Multiple Qubits} 
\label{sec:ex2}
\begin{figure*}
    \centering
    \scalebox{0.75}{
    \includegraphics[width=\textwidth]{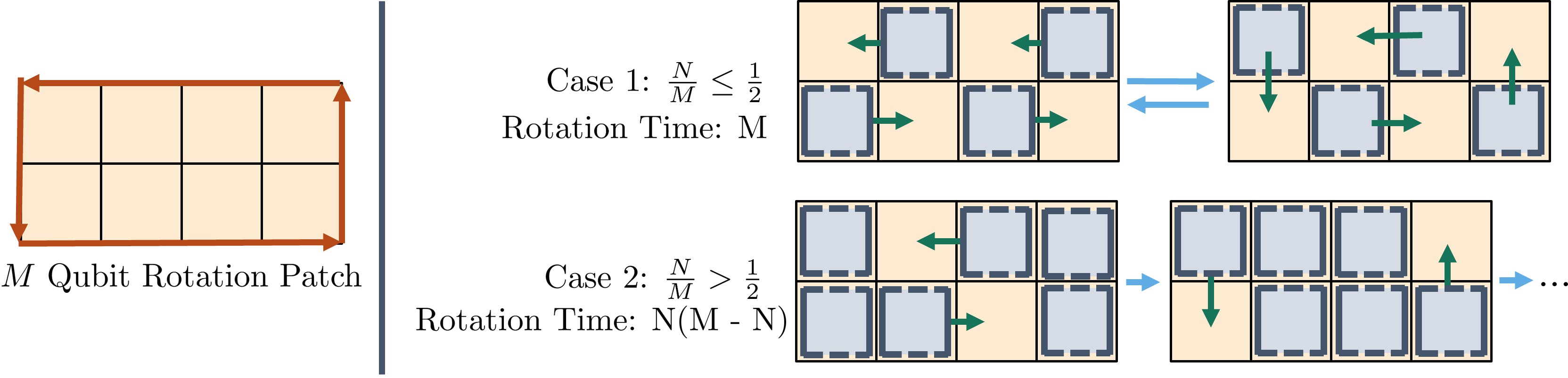}}
    \caption{The generalized case from before. On the left we give an abstract representation of a rotational patch where qubits can move through. On the right we give two examples for different occupancy, i.e. the number of qubits mapped to a given rotation unit. First is when occupancy is less than 1/2 in which case every qubit can move every cycle since they are never blocked. The second is when occupancy is high and therefore some qubits can move. In both cases every qubit can still be guaranteed an equal amount of time in every position but the total rotation duration is much longer is the second case.}
    \label{fig:cases}
\end{figure*}
As a first generalization, consider the case of a single logical qubit $\ket{A}$ mapped into a grid of $M$ positions. The previous example is simple to extend such that the probability of success is given as
\begin{align}\label{eq:avgcase}
    \prod_{i = 1}^M (1 - p_L(q_i))^\frac{T}{M}.
\end{align}
This example is fairly contrived since we can better mitigate variable error rates by expanding the code distance of $\ket{A}$ given the available hardware space. In reality, we expect much less availability and this motivates the most general example of having $N$ logical qubits $\ket{A_1}, ..., \ket{A_N}$ mapped into $M$ hardware positions $q_1, ..., q_M$, with $M > N$, as in Figure \ref{fig:cases}. Ideally, we can ensure that each of the $N$ qubits occupies every position for an equal amount of time, though in practice as we will see this is not easy because of ``rotation stalls'' caused by multi-qubit operations. For now, consider memory only. We can consider the \textit{occupancy} of the block of $M$ positions to be $o = N/M$. We can break it down into two cases as in Figure \ref{fig:cases}: 1. $o \le 1/2$ in which case all data qubits can move every cycle. It will then take $M$ cycles for every qubit to occupy every position. 2. $o > 1/2$ in which case only $M - N$ of the data qubits can move and the others are stalled. In this limit, the occupancy can still be made equal for all qubits in all positions. Therefore, regardless of the case, the estimated success rate for \textit{each} qubit is as in Eq. \ref{eq:avgcase} for sufficiently large $T$.

We should compare the physical qubit requirement of our proposal versus one which increases the distance $d \rightarrow d + 1$. A distance $d$ surface code requires $f(d) = 2d^2 - 1$ (data plus ancilla). Then we can support adding additional ancilla $m = M - N$ when $(N + m)f(d) \le Nf(d+1)$. E.g. when $d =3$ then $m \le N$, i.e. $o \approx 50\%$ though when $d=9$ then $m < N/3$ so $o \approx 66\%$ so our strategy becomes less powerful as the number of available hardware qubits increases, for which we also similarly expect hardware variance to decrease. However, we anticipate some positive compilation side effects which can reduce the total computation time as well.

%% file: content/60_conclusion.tex
\section{Conclusion}
NISQ-era devices suffer from multiple sources of error. QEC is used to mitigate these sources of error. The surface code, a topological quantum error correcting code, is ubiquitous throughout current literature for its efficacy as a quantum error correcting code. Logical qubits can be encoded into surface codes, and are mapped onto a board of tiles, wherein each tile is at least $d\times d$. Logical qubits may interact and maneuver around the board through either lattice surgery or braiding.

In this short paper, we explored the idea of cyclic lattice surgery for mitigating hardware heterogeneity. We started by demonstarting the uncertainty caused by spatial and temporal variation in logical error rates. Following this, we proposed how dynamic remapping could mitigate sources of error in quantum systems, such as burst errors and drift. This is done via the enforcement of an average logical error rate on each qubit. Finally, we put forth a dynamic mapping compilation framework, CQM, which we anticipate can dynamically remaps qubits with minimal time and resource overheads.